# RADIATION ENVIRONMENT IN EARTH-MOON SPACE: RESULTS FROM RADOM EXPERIMENT ONBOARD CHANDRAYAAN-1

S. V. VADAWALE[†], J. N. GOSWAMI

*Physical Research Laboratory, Ahmedabad - 380009, India*

T. P. DACHEV, B. T. TOMOV

*Solar-Terrestrial Influences Institute, Bulgarian Academy of Science, Sofia, Bulgaria*

V. GIRISH

*ISRO Satellite Center, Bangalore - 000000, India*

The Radiation Monitor (RADOM) payload is a miniature dosimeter-spectrometer onboard Chandrayaan-1 mission for monitoring the local radiation environment in near-Earth space and in lunar space. RADOM measured the total absorbed dose and spectrum of the deposited energy from high energy particles in near-Earth space, en-route and in lunar orbit. RADOM was the first experiment to be switched on soon after the launch of Chandrayaan-1 and was operational till the end of the mission. This paper summarizes the observations carried out by RADOM during the entire life time of the Chandrayaan-1 mission and some the salient results.

## 1. Introduction

The near Earth space is populated with energetic particles of galactic, solar and trapped origin with energies spread over keV/amu to GeV/amu. Further, the radiation environment is affected by the solar activity and hence is highly variable in time. It is necessary to have as much continuous and thorough monitoring of this radiation environment as possible when a spacecraft is exploring the near Earth space or beyond, e.g. exploration of other satellites or planets in the solar system. Such monitoring provides input to asses the radiation hazard, particularly for electronic subsystems of the spacecraft and payloads on board and also help us in having a better understanding of the space radiation environment around the Earth and en route and orbiting a specific target, e.g., the Moon in the present case. The RADOM experiment was selected from the AO (Announcement of Opportunity) proposals received for India's first mission to the Moon – Chandrayaan-1 [1,2]. It consists of a small semi-conductor (Si-PIN) radiation detector, which measures the deposited energy by high energy particles incident on it, provide the total absorbed dose and also generate spectrum of the deposited energy. In this paper, we summarize the observations carried out by the RADOM experiment, starting form the early phase of the Chandrayaan-1 launch through the end of the mission.

---

[†] Corresponding author (email: *santoshv@prl.res.in*)

## 2. Space Radiation Environment

The Space Radiation Environment is constituted of three basic components; Galactic cosmic rays; solar energetic particles and the trapped radiation belts around Earth.

### 2.1. *Galactic Cosmic Rays (GCR)*

The Galactic Cosmic Rays are high energy particles coming from outside the solar system and are modulated by solar activity within the heliosphere. The energy spectra of GCR in the inner solar system effectively start at about 100 MeV/amu and extend beyond GeV/n. GCRs arrive isotropically through out the space and consist of ~2% electrons and ~98% hadrons. Among the hadrons, 90% are protons, 9% are alpha particles and the rest are nuclei of higher Z elements such as C, O, Fe etc. At energies greater then ~$10^{12}$ eV their flux is very small and it is the lower energy end of the spectrum where the GCR particles posses the real radiation hazard. Up to 1 GeV, the flux and spectra of GCR particles is strongly influenced by the solar activity and hence shows modulation which is anti-correlated with solar activity. For further information on GCR the reader is requested to consult standard textbooks on the subjects such as by Ginzburg & Syrovatskii [3] and Gaisser [4] or recent reviews such as by Cronin [5] and Strong et al.[6]

### 2.2. *Solar Energetic Particles (SEP)*

Solar Energetic Particles (SEP) with energies primarily in the 1-100 Mev/amu region originate from energetic solar flares and coronal mass ejection events[7]. Energy of SEPs can extend up to 1 GeV and their intensity can rise as high as $10^4$ particle $cm^{-2}$ $s^{-1}$ $sr^{-1}$ during a large SEP event [8]. Since SEPs are highly variable and unpredictable events, they pose major radiation hazard for a space craft or its subsystems during space exploration. Therefore it is essential to have a real time radiation monitoring system in such missions.

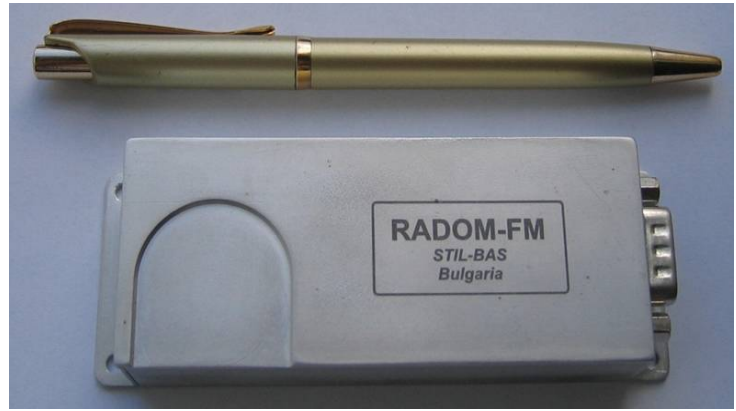

Figure 1. Flight model of the RADOM instrument.

### 2.3. *Trapped Radiation Belts*

Radiation Belts are regions of high concentration of energetic electrons and protons trapped within the Earth's magnetosphere. There are two distinct belts of toroidal shape surrounding the Earth where the energetic charged particle gets trapped in the Earth's magnetic field. The inner radiation belt, located between about 0.1 to 2 Earth radii, consists of electrons with energies up to 10 MeV as well as protons with energies up to ~100 MeV. The outer radiation belt starts from about 4 Earth radii and extends to about 9-10 Earth radii in the anti-sun direction. The outer belt mostly consists of electrons whose energy and concentration are much more variable then in the inner belt. For further information on the radiation belts the reader can consult an excellent textbook by M. Walt [9].

### 3. RADOM onboard Chandrayaan-1

The RADOM is a very small (~10 x 5 x 2 cm$^3$) and low-weight (~100 gm) instrument (see Figure 1). It consists of a single 0.3 mm thick Si-PIN semiconductor detector with 2 cm$^2$ area; one low noise hybrid charge-sensitive preamplifier (A225F from Amptek Inc.); a fast 12 bit ADC; 2 microcontrollers and buffer memory. Pulse height analysis technique is applied for determination of the energy deposited during individual particle interaction. The measurement of energy deposited by individual particles can then be used to infer the absorbed dose in the silicon detector. The unit is managed by microcontrollers through specially developed firmware. RS232 interface facilitates transmission of data stored in the buffer memory to the Chandrayaan-1 telemetry. A more detailed description of the RADOM instrument is given by Dachev et al. [10]. RADOM records the spectrum of the deposited energy in 256 channels with a pre-defined exposure time (typically 30 s) for a given spectrum. The exposure time can be changed by ground command, if necessary. After finishing a measurement cycle the data are stored in the buffer memory and the accumulated data are transmitted through RS232 interface for telemetry.

The absorbed dose in Grays is calculated from the measured spectrum as:

$$D = K \sum_{i=1}^{256} \frac{A_i k_i}{MD}$$

Where, $A_i$ is the counts in $i^{th}$ channel and MD is mass of the sensitive part of the detector (in kg). K and $k_i$ are coefficients determined form the calibration of the instrument. RADOM is very similar to some of the earlier instruments such as the Liulin-E094 Mobile dosimetry units [11,12] and R3DE/R3DR dosimeters flown onboard the International Space Station (ISS) [13].

## 4. Observations

### 4.1. *Initial Earth orbits*

RADOM was the first scientific payload to be switched on after the launch of the Chandrayaan-1. Measurements started on 22nd October 2008 two hours after the launch, when the spacecraft was orbiting the Earth in a low elliptical orbit with apogee of about 23000 km and perigee of about 250 km. Preliminary results of RADOM observations during the first Earth orbits of Chandrayaan-1 were presented by Dachev et al. [14]. Data obtained by RADOM during later Earth orbits of Chandrayaan-1 are shown in Figure 2 in terms of variation of total particle flux (top panel), total absorbed dose (second panel), gray scale spectrum of deposited energy (third panel) and the spacecraft altitude (bottom panel) with time from launch. It can be seen that the variation of total particle flux and deposited dose during the perigee of these orbits is almost same as that presented by Dachev et al. [14]. The variation of the deposited energy spectrum clearly distinguishes the outer and the inner radiation belts, slot region between them as well as almost radiation free region below the inner radiation belt. The observed total particle flux in the outer radiation belt is ~1.5×10$^4$ particle cm$^{-2}$ s$^{-1}$, which is higher compared to the flux observed in the inner radiation belt, ~9×10$^3$ particle cm$^{-2}$ s$^{-1}$. However, the total absorbed dose rate in the outer radiation belt is ~4×10$^4$ µGy h$^{-1}$, which is much lower compared to the total dose rate in the inner radiation belt, ~1.3×10⁴ μGy h⁻¹. This shows that the energy deposited by individual particles in the outer belt is much lower compared to that in the inner belt, which in turn implies that the particle population in the outer belt is dominated by electrons where as particle population in the inner belt is dominated by heavy particles (mainly protons).

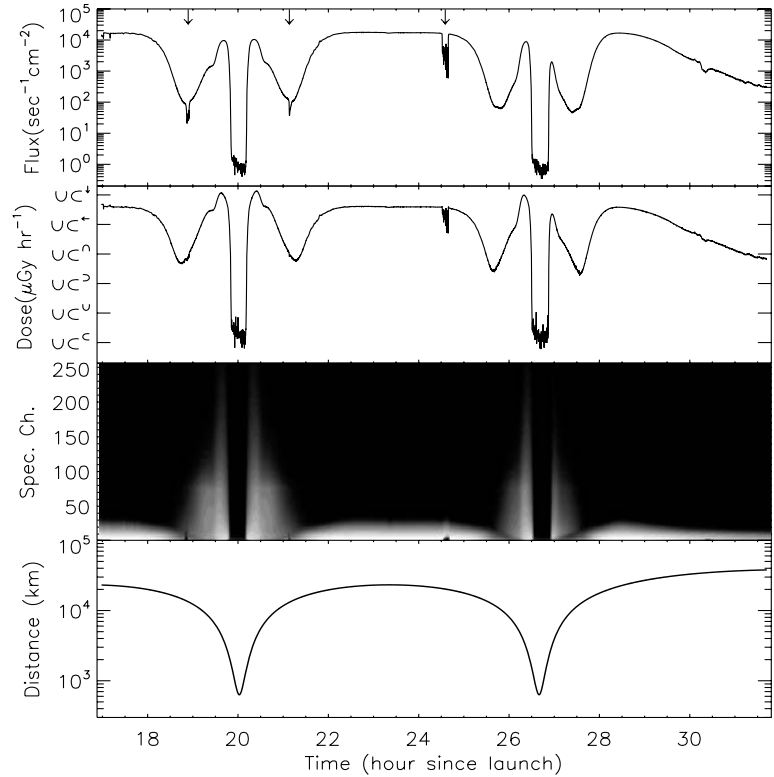

Figure 2. RADOM observations during initial earth orbits of Chandrayaan-1. Top two panels show the particle flux and corresponding absorbed dose rate (in μGy h⁻¹, see text). Third panel show the 256 channel spectrum of deposited energy. The 256 channels are on vertical axis and the gray scale represents log of counts in each channel. The bottom panel shows spacecraft altitude from earth's surface. Some discontinuities in the flux and dose rate plots (marked by arrow) are due to the data corruption.

The first orbit raising maneuver for Chandrayaan-1, increasing apogee to ~40000 km, was carried out during the second perigee passage 27 hours after the launch. This is evident from the decreasing particle flux (Figure 2) as the spacecraft passes through the outer belt. It can be seen that the total flux and

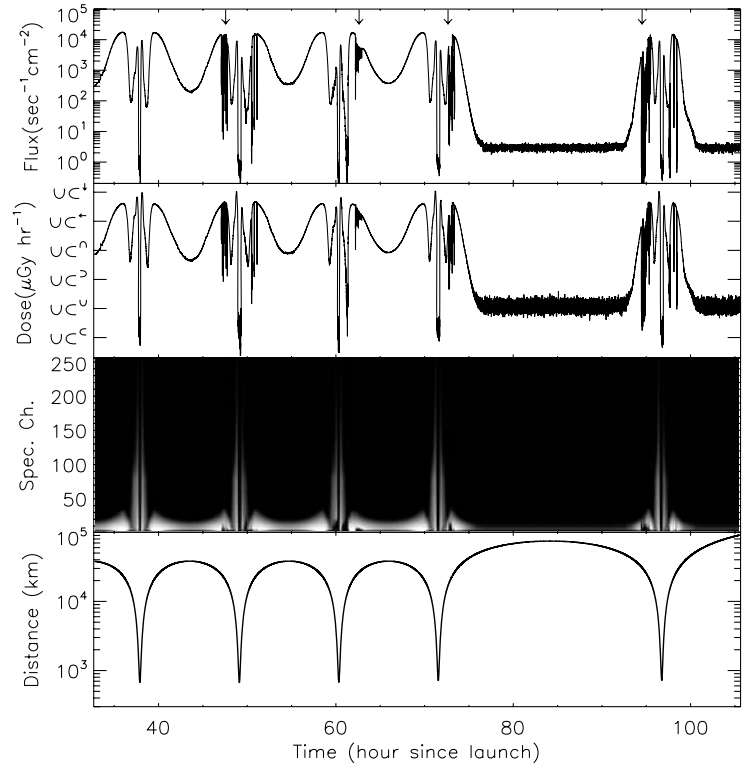

Figure 3. RADOM observations during initial earth orbits of Chandrayaan-1. Panels are same as in Figure 2. Discontinuities in the flux and dose rate plots (marked by arrow) are due to the data corruption

deposited energy spectra are asymmetric on the two sides of the perigee, a result of the outgoing spacecraft trajectory passing through the periphery of the inner radiation belt. Figure 3 shows the variation of the same quantities during some of the later orbits with apogee of ~37000 km and ~75000 km. It can be seen that the maximum flux and dose rate as well as their variations are consistent during each perigee passage. At the apogee of ~37000 km, particle flux is lower then the maximum observed in the outer belt which indicates that the apogee is close to the outer edge of the belt. Orbits with apogee of ~75000 km clearly shows the boundary region of the outer radiation belts. These observations shows that the inner radiation belt starts at ~1500 km from the Earth's surface and extends up to ~10000 km whereas the outer radiation belt starts from

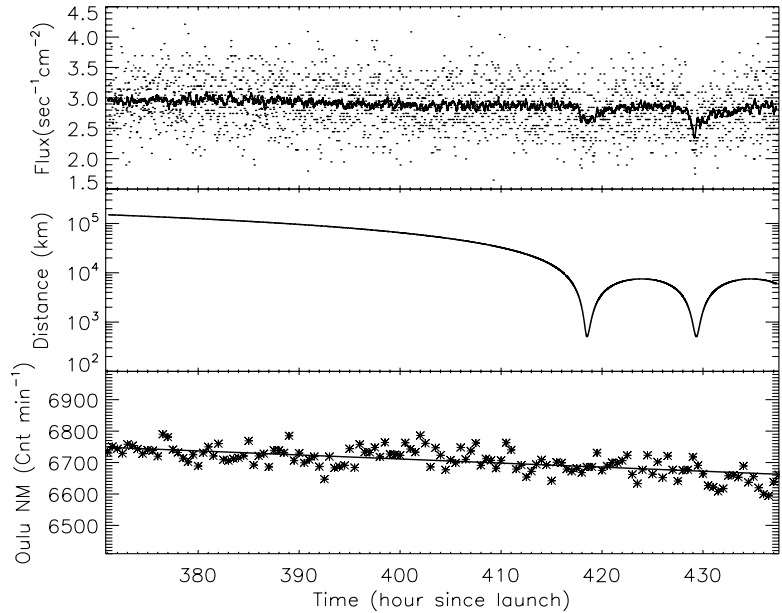

Figure 4. RADOM observations during lunar transfer trajectory and lunar orbit capture. The distance is from the moon. The particle flux is almost constant; however, there is slight decreasing trend. Comparison with Oulu neutron monitor data shows that this decreasing trend is real.

~15000 km and extends up to ~53000 km. Maximum particle concentration in the inner belt is observed at ~6000 km whereas the maximum particle concentration in the outer belt is observed at ~25000 km.

## 4.2. *Entry into deep space and lunar orbit capture*

Chandrayaan-1 was placed in the lunar transfer trajectory on 13th day after launch (3rd November) and lunar orbit capture maneuver was carried out on 18th day after the launch (8th November). Figure 4 shows RADOM observations for about 2 days before the lunar orbit capture and about one day after it. The top panel of this figure shows the measured total particle flux and the second panel shows the distance from the moon. The measured total particle flux is ~3 particles cm$^{-2}$ s$^{-1}$, and corresponding absorbed dose rate is ~12 μGy h$^{-1}$. It can be seen that the flux is almost constant except for the slight decrease during periselene which indicates enhanced shielding of the cosmic rays by the moon itself as the spacecraft approaches the moon. However a closer look at the top panel reveals marginal decreasing trend in the average particle flux. In particular, at distances far away from the moon, the spacecraft is in the deep space so the decrease can not be explained by any shielding effect and suggests that there could be intrinsic decrease in the cosmic-ray intensity. Independent observations by the Oulu neutron monitor (shown in the bottom panel) indeed confirms that the cosmic-ray intensity showed a slight decreasing trend during this period, which could be related to the solar activity.

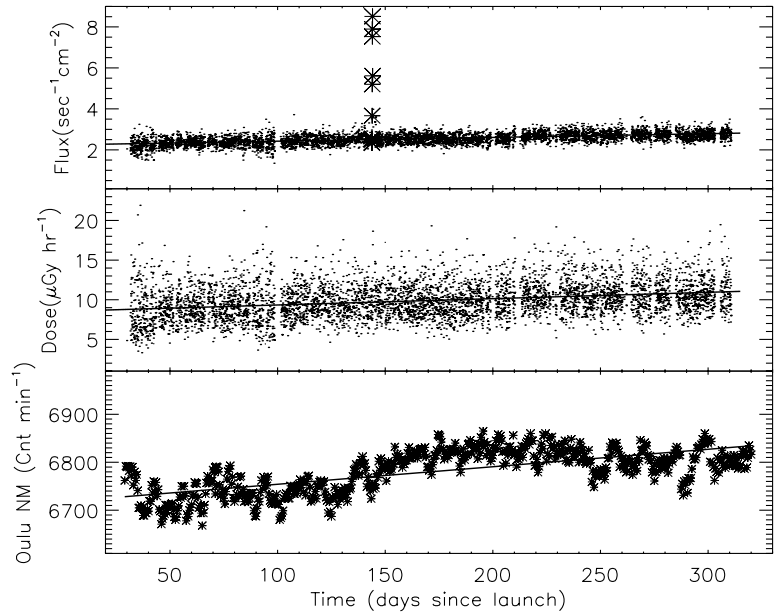

Figure 5. Long term monitoring of the lunar radiation environment. The particle flux (top panel) and dose rate (second panel) are almost constant, except a minor SEP event on 144[th] day from launch. However there is slight increasing trend in flux. Comparison with the Oulu newtron monitor data (third panel) shows that this is a real increase in the GCR intensity.

### 4.3. *Lunar radiation environment*

Chandrayaan-1 was placed in the final circular polar lunar orbit at 100 km altitude on 14[th] November 2008. Figure 5 shows the RADOM observations in the lunar orbit till the end of the mission. It can be seen that particle flux is almost constant throughout the duration of the mission except a small flare around 144[th] day (15[th] March 2009). However, a closer look reveals a marginal increasing trend in the flux. Again, comparing with the independent measurements of cosmic ray intensity by Oulu neutron monitor data shows that this is a real increase in the cosmic ray intensity due to decreasing effect of solar magnetic fields. Overall particle flux in the 100 km orbit was found to be ~2.5 particles cm$^{-2}$ s$^{-1}$, and the corresponding absorbed dose rate was ~10 μGy h$^{-1}$. During the last three months of the mission, Chandrayaan-1 was in 200 km orbit, where the flux and dose rate increased slightly to ~2.8 particles cm$^{-2}$ s$^{-1}$ and ~11 μGy h$^{-1}$ respectively. The solid angle of acceptance for open space at 200 km altitude increases by about 10 % then that at 100 km altitude. The observed increase of particle flux at 200 km can thus be explained as due to reduced self-shielding of GCR by the moon.

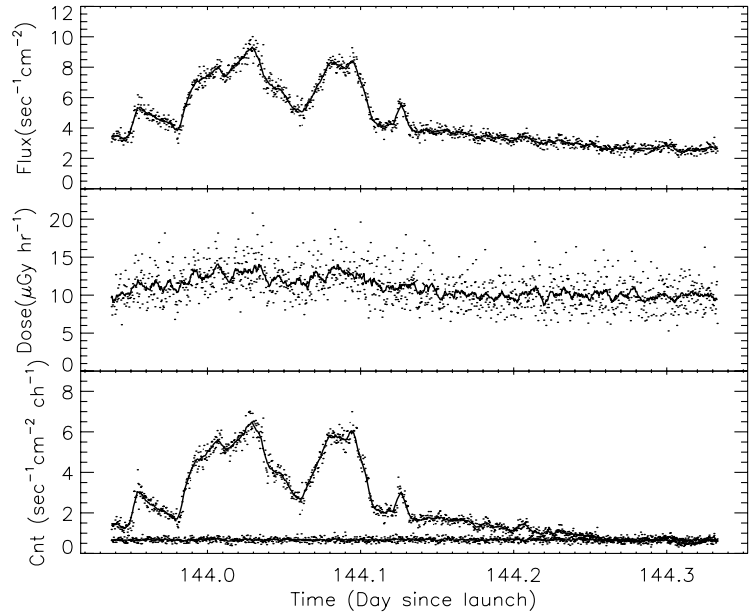

Figure 6. RADOM observation of an SEP event. The total particle flux (top panel) is varying significantly, however, corresponding absorbed dose rate (second panel) is not changing much. Bottom panel shows variation in the first channel (upper trace) and second channel (lower trace) of the 256 spectrum. This shows that the flux variation is only observed in first channel.

The observation of flare on 15th March 2009 is an interesting case. Figure 6 shows the data zooming around the time of this flare occurrence. The particle flux increases significantly during the flare; however the absorbed dose is almost constant. The spectrum of deposited energy during this flare shows that the enhancement is only in the first channel of the spectrum, which means that the increase in flux is only due to increase of electron concentration. The GOES solar monitoring data during this period also shows increase in electron and proton flux. However, the detailed inspection of the GOES data indicates that the proton flux enhancement is limited to energies <10 MeV, whereas the electron flux enhancement extends beyond energies of 2 MeV. Since the lower energy cutoff for electron and proton detection by RADOM is ~0.8 MeV and ~17 MeV respectively [10], RADOM could not detect increase in the low energy proton flux and only the increase in electron flux was detected. This shows that apart from detecting the radiation belt particles and the galactic cosmic rays, RADOM could also detect solar energetic particle events. It should be noted that the solar activity was at the historically lowest during the operation of the Chandrayaan-1, and hence only one solar event is detected by RADOM. There were some small solar flares, a particularly noticeable one on 6[th] July 2009 that was detected by the X-ray Solar Monitor (XSM) experiments onboard Chandrayaan-1 [15], during the operation of Chandaryaan-1, however RADOM did not detect any significant enhancement in particle flux during these flares. Directional ejection of the solar energetic particles could be the reason for this.

## 5. Summary

RADOM observations beginning within two hours after the launch of the Chandrayaan-1 and continuing till the end of the mission demonstrated that it could successfully characterize different radiation fields in the Earth and Moon environments. Signature and intensity of proton and electron radiation belts, solar energetic particles as well as galactic cosmic rays were well recognized and measured. Effect of solar modulation of galactic cosmic rays could also be discerned in the data. The electron radiation belt doses reached ~40000 μGy/h, while the maximum flux recorded was ~15000 particle cm$^{-2}$ s$^{-1}$. The proton radiation belt doses reached the highest values of ~130000 μGy/h, while the maximum flux was ~9600 particle cm$^{-2}$ s$^{-1}$. Comparison of these results with other similar instruments on board ISS shows good consistency, indicating nominal performance RADOM. Outside the radiation belts, en-route to the Moon, the particle flux (~3 particle cm$^{-2}$ s$^{-1}$) and corresponding dose were very small (~12 μGy) which further decreased slightly in the lunar orbit because of the shielding effect of the Moon. Average flux and dose in lunar orbit were ~2.5 particle cm$^{-2}$ s$^{-1}$ and ~10 μGy h$^{-1}$ respectively at 100 km orbit. These increased to ~2.8 particle cm$^{-2}$ s$^{-1}$ and ~11 μGy h$^{-1}$ respectively, at 200 km orbit.

The total accumulated dose during the transfer from Earth to Moon was found to ~1.3 Gy. Due to the lack of significant solar activity only minor variations in the particle flux and dose were observed in the lunar orbit. Comparison of the RADOM observations with theoretical models of radiation environment of both the moon and the Earth are in progress; some of the preliminary results were presented elsewhere [16, 17].

### Acknowledgments

The authors would like to thank the entire Chandrayaan-1 team whose dedicated efforts made it a highly successful mission.

### References

1. J. N. Goswami and M. Annadurai, *Curr. Sci.*, **96**, 486 (2009)
2. J. N. Goswami *41st LPSC*, id.1591 (2010)
3. V. L. Ginzburg & S. I. Syrovatskii, *"The Origin of Cosmic Rays"*, Macmillan (1964)
4. T. K. Gaisser, *"Cosmic Rays and Particle Physics"*, Cambridge Uni. Press (1990)
5. J. W. Cronin, *Rev. Mod. Phy.* **71**, 165 (1999)
6. A. W. Strong, *Annu. Rev. Nucl. Part. Sci.*, **57**, 285 (2007)
7. D. V. Reames, *Space Sci. Rev.*, **85**, 327 (1998)
8. H. Hudson & J. Ryan, *Annu. Rev. Astron. Astrophys.*, **33**, 239 (1995)
9. M. Walt. *"Introduction to Geomagnetically Trapped Radiation"*, Cambridge. Uni. Press (2005)
10. T. P. Dachev et al., *Curr. Sci.*, **96**, 544 (2009)
11. T. P. Dachev et al., *Adv. Spece Res.*, **30**, 917 (2002)
12. J. W. Wilson et al., *Adv. Spece Res.*, **40**, 1562 (2007)
13. T. Dachev et al., *Proc. 11th International Science Conference on Solar-Terrestrial influences*, Sofia, p. 171 (2005)
14. T. P. Dachev et al., *40th LPSC*, id. 1274 (2009)
15. S. Narendranath et al., *41st LPSC*, id. 1882 (2010)
16. G. de Angelis et al., *40th LPSC*, id. 1310 (2009)
17. G. de Angelis et al., *41st LPSC*, id. 1711 (2010)